\documentclass[final,3p,times,sort&compress]{elsarticle}

\usepackage{multirow,setspace,times,amssymb,amsmath,graphicx,color,rotating,subfigure,url}

\usepackage{lineno}
\usepackage{booktabs}%插入表格线加粗合并
\usepackage{threeparttable}   %给表格加注释
\begin{document}

\begin{frontmatter}

\title{Profitability of simple technical trading rules of Chinese stock exchange indexes}　

\author[BS]{Hong Zhu}
\author[BS,RCE]{Zhi-Qiang Jiang\corref{cor}}
\ead{zqjiang@ecust.edu.cn}
\author[RCE,AS]{Sai-Ping Li}
\author[BS,RCE,SS]{Wei-Xing Zhou\corref{cor}}
\ead{wxzhou@ecust.edu.cn} %
\cortext[cor]{Corresponding authors.}% Address: 130 Meilong Road, P.O. Box 114, School of Business, East China University of Science and Technology, Shanghai 200237, China, Phone: +86 21 64253634, Fax: +86 21 64253152.}

\address[BS]{Department of Finance, School of Business, East China University of Science and Technology, Shanghai 200237, China}
\address[RCE]{Research Center for Econophysics, East China University of Science and Technology, Shanghai 200237, China}
\address[AS]{Institute of Physics, Academia Sinica, Nankang, Taipei 11529, Taiwan}
\address[SS]{Department of Mathematics, School of Science, East China University of Science and Technology, Shanghai 200237, China}

\begin{abstract}
  Although technical trading rules have been widely used by practitioners in financial markets, their profitability still remains controversial. We here investigate the profitability of moving average (MA) and trading range break (TRB) rules by using the Shanghai Stock Exchange Composite Index (SHCI) from May 21, 1992 through December 31, 2013 and Shenzhen Stock Exchange Composite Index (SZCI) from April 3, 1991 through December 31, 2013. The $t$-test is adopted to check whether the mean returns which are conditioned on the trading signals are significantly different from unconditioned returns and whether the mean returns conditioned on the buy signals are significantly different from the mean returns conditioned on the sell signals. We find that TRB rules outperform MA rules and short-term variable moving average (VMA) rules outperform long-term VMA rules. By applying White's Reality Check test and accounting for the data snooping effects, we find that the best trading rule outperforms the buy-and-hold strategy when transaction costs are not taken into consideration. Once transaction costs are included, trading profits will be eliminated completely. Our analysis suggests that simple trading rules like MA and TRB cannot beat the standard buy-and-hold strategy for the Chinese stock exchange indexes.
\end{abstract}


\end{frontmatter}

\section{Introduction}

Investors investigate market behavior by using technical analysis with the aim to predict future market trends. From the microscopic view, investors may get profits by applying technical analyses to investment decision-making. From the macroscopic view, the significant profitability of trading rules is often interpreted as evidence against market efficiency. The debates on the usefulness of technical analysis have attracted considerable research interests in recent years. Most of the early studies support the random walk hypothesis, which means that technical analyses are invalid. The traditional statistical tests have been applied to demonstrate the failure of technical analyses \cite{Fama-Blume-1965-JB,Levy-1967-FAJ}. On the other hand, most of the studies after 1992 provide strong arguments that technical trading rules can forecast the market trend and earn excess returns. Brock, Lalonishok, and Lebaron employed the $t$-tests to check whether the returns conditioned on the trading signals generated by the MA and TRB rules were significantly different from unconditioned returns \cite{Brock-Lalonishok-Lebaron-1992-JF} and found that the technical trading rules had significant predictive ability in U.S. market during the period from 1897 to 1986.

By applying the same methods in the study of Ref.~\cite{Brock-Lalonishok-Lebaron-1992-JF}, many scholars advocate that the technical trading rules are profitable in different markets. Bessembinder and Chan found that the technical trading rules are successful in predicting stock price movements in emerging markets such as Malaysia, Thailand and Taiwan, and the forecast ability are greatly reduced in relatively developed Hong Kong and Japan markets \cite{Bessembinder-Chan-1995-PBFJ}. Even if the transaction costs and nonsynchronous trading are taken into account, the possibility of trading profits in Malaysia, Thailand and Taiwan cannot be dismissed in their study periods. Hudson, Dempsey, and Keasey also found that the MA rules coul provide trading profits when they are applied on the Financial Times Industrial Ordinary Index from 1935 to 1994. However, these trading profits would be eliminated by the inclusion of transaction costs \cite{Hudson-Dempsey-Keasey-1996-JBF}. Ito evaluated the profitability of technical trading rules in Pacific-Basin equity markets from 1980 to 1996 and found that trading rules had the predict ability in Japan, Canada, Indonesia, Mexico, and Taiwan markets, not in US market \cite{Ito-1999-PBFJ}. Lai and Balachandher focussed on the predictability of VMA and FMA rules on Kuara Lumpur Stock Exchange Composite Index which covered the period from 1977 to 1999 in the Malaysian Stock Market and found that VMA and FMA rules could outperform the buy-and-hold strategy even with transaction costs \cite{Lai-Balachandher-Nor-2002-QJBE} . Vasiliou, Eriotis, and Papathanasiou applied MA and MACD strategies to the Athens General Index from the beginning of 1990 till the end of 2004 and find that MA strategies (annual return 36.10\%) and MACD strategies (annual return 55.65\%) was able to outperform the buy-and-hold strategy (annual return 12\%) in Athens Stock Market \cite{Vasiliou-Eriotis-Papathanasiou-2006-OR}. Mitra and Choe et al found the strong arguments of the usefulness of technical trading rules in Indian stock market and G-7 stock markets (Canada, France, Germany, Italy, Japan, United Kingdom and United States) \cite{Mitra-2011-QF, Choe-Krausz-Nam-2011-RQFA}. Yu et al investigated five south-east markets (Singapore, Malaysia, Thailand, Indonesia and the Philippines) by means of 60 kinds of VMA, FMA and TRB trading rules during the period from 1991 to 2008 and found quite similar results as in Ref.~\cite{Bessembinder-Chan-1995-PBFJ}. Trading rules have stronger predictive power in the emerging stock markets of Malaysia, Thailand, Indonesia, and the Philippines than in the more developed stock market of Singapore \cite{Yu-Nartea-Gan-Yao-2013-IREF}. Coe and Laosethakul tested four technical trading rules (the arithmetic moving average, the relative strength index, a stochastic oscillator and its moving average) against 576 stocks which also included S\&P 100, the NASDAQ 100 and the S\&P Midcap 400 indices and found that none of these technical trading rules could outperform the market. In Chinese markets, it was found that the technical trading rules could bring excess returns \cite{Lin-Zeng-Tang-2000-cnJUESTC,Sun-Fang-2004-cnJSUFE,Sun-2005-cnJQTE,Wang-Zeng-Li-2007-cnJQTE} and the transaction costs had effects on the overall performance of trading rules \cite{Tang-Zeng-Tang-2002-cnJIEEM,Dai-Wu-2002-cnJQTE}.

When a set of data has been used for many times to inference and model selection, data snooping effects will occur. Hence, some scholars doubt the positive evidence of the profitability of technical trading rules because of the data snooping effect. White put forward a White's Reality Check (WRC) method to test whether a financial market trading strategy generate returns superior to the benchmark by considering the effect of data snooping \cite{White-2000-Em}. Sullivan et al found that the results of \cite{Brock-Lalonishok-Lebaron-1992-JF} passed the WRC tests.  Li found that the technical trading rules could not predict the future trend very well for Hushen 300 index \cite{Li-2010-cnSIF} by means of WRC tests.

In this paper, we will apply technical trading rules to Shanghai and Shenzhen markets and to examine whether any of these technical trading rules would generate higher profits against the buy-and-hold strategy by means of $t$-test and WRC tests. The rest of the paper is arranged as follows. Section 2 discusses the technical trading rules. The data and methodology of $t$-test and WRC test are presented in Section 3. Section 4 describes the empirical results and Section 5 is the conclusion.

\section{Trading rules}
\subsection{Moving averages}

According to the moving average (MA) rules, which is one of the most popular trading rules in technical analysis, trading signals are triggered if the short term moving average penetrates the long term moving average. More specifically,  if the short term moving average rises above (or falls below) the long term moving average, a buy signal (or sell signal) will be generated.  One can however observe many intersections (markers of trading signals) between both moving averages in the range-bound market. These are ``fake'' signals, which hardly provide any profits, but would instead increase the transaction costs.  In order to alleviate this situation,  we impose a criterion that the short term average must be greater (less) than the long term average by a predefined percentage band before a trading signal is generated.  For comparison, we test both MA rules with and without such a band.  Once the trading signal is generated, the position will be changed according to the following rules: the variable-length moving average (VMA) rule and the fixed-length moving average (FMA) rule. VMA rules require investors to hold the position until the condition that generates the former signal is no longer valid. FMA rules call for investors to hold the position for a fixed number of days, during which all the new signals are ignored.

The MA price on day $t$ with an averaging window size $n$ is defined as follows,
\begin{equation} \label{Eq:MA}
  m(t, n)=\frac{1}{n} \sum_{i=0}^{n-1} p(t-i)  \,\,\, .
\end{equation}
The position states generated by the VMA rules can be formulated as,
\begin{eqnarray}\label{Eq:VMASignal}
{\rm{long~position}} & m(t, n_s)>(1+b) m(t, n_l)\\
{\rm{short~position}} & m(t, n_s)<(1-b) m(t, n_l)\\
{\rm{closed~position}} & (1+b) m(t, n_l) \ge m(t, n_s) \ge (1-b) m(t, n_l)
\end{eqnarray}
where $n_s$ is the window size of short averages, $n_l$ is the window size of long averages, and $b$ is the band. We write the parameters of the VMA rules as $(n_s, n_l, b)$, where $n_s$ is chosen from 1, 2, and 5; $n_l$ is chosen from 20, 50, 150, and 200; and $b$ is chosen from 0 and 0.01. This leads to 24 VMA rules.

The trading signals generated by the FMA rules can be formulated as,
\begin{eqnarray}\label{Eq:FMASignal}
{\rm{buy}} & m(t-1, n_s)<(1+b) m(t-1, n_l) ~~{\rm{and}} ~~m(t, n_s)>(1+b) m(t, n_l)\\
{\rm{sell}} & m(t-1, n_s)>(1-b) m(t-1, n_l) ~~{\rm{and}}~~m(t, n_s)<(1-b) m(t, n_l)
\end{eqnarray}
The position will be closed after being held for $C$ days, which means that the number of long (short) position days equals to the number of buy (sell) signals multiplied by $C$. Hence, we have the set of parameters $(n_s, n_l, b, C)$. The chosen values of $n_s$ and $b$ for the FMA rules are the same as those for the VMA rules. The values of $n_l$ can be 50, 150, and 200. $C$ is fixed at 10. These parameters result in 18 FMA rules.

\subsection{Trading range break-out rules}

Trading range break-out (TRB) rules are known as support and resistance rules. In our tests, the support and resistance levels are defined as the minimum and maximum prices over the previous 50, 150, and 200 days respectively. A buy (sell) signal is generated when the price penetrates the resistance (support) level. In order to reduce noisy signals, we also use additional bands of 0 and 0.01 to generate trading signals.  These lead to 6 TRB rules. Once the trading signal is activated, the position will be held for $C$ days, during which all the trading signals are ignored. Analogous to the FMA rules, we also set $C=10$ days.

The trading signals generated by the TRB rules can be formulated as,
\begin{eqnarray}\label{Eq:TRB}
{\rm{buy}} & p(t-1)  < (1+b) p_{\max}~~{\rm{and}}~~p(t)  > (1+b) p_{\max}\\
{\rm{short}} & p(t-1)  > (1-b) p_{\min}~~{\rm{and}}~~p(t)  < (1-b) p_{\min}
\end{eqnarray}
where $p_{\max}$ and $p_{\min}$ are the local maximum and minimum values for the previous 50, 150, and 200 days.

\section{Data and Methodology}
\subsection{Data}
The profitability of our technical trading rules are evaluated on two famous indexes (SHCI and SZCI) in Chinese stock market. We download both daily indexes from the financial data provider RESSET. Both indexes cover a period from April 3, 1991 through December 31, 2013, during which there were a total of 5593 trading days. The daily returns, defined as the log differences of the daily prices are calculated for each index. We find that the return on May 21, 1992 was about 105\% for SHCI, which was caused by the cancellation of price limit for 15 stocks in Shanghai stock market. In order to avoid the influence of this big return artifect on the performance of our trading rules, we discard the data before May 21, 1992 for SHCI, which leads to 5276 data points in total.  Table~\ref{return_test} lists the basic statistics of daily returns for both indexes. When compared to a normal distribution, both return distributions have excess kurtosis and right skewness.
% Table generated by Excel2LaTeX from sheet 'Index Result'
\begin{table}[!htp]
  \centering
  \small
  \caption{Basic statistics of daily returns for SHCI and SZCI.}
    \begin{tabular}{ccccc}
    \toprule
          & SHCI 1-day & SZCI 1-day \\
    \midrule
    mean  & 0.00097  & 0.00042  \\
    std   & 0.023  & 0.022  \\
    skew  & 1.186  & 0.553  \\
    kurtosis & 22  & 17  \\
    Obsevation & 5275  & 5592  \\
    \bottomrule
    \end{tabular}%
  \label{return_test}%
\end{table}%

\subsection{Methodology}
\subsubsection{Traditional T-test Method}
We employ the same method as in Ref.~\cite{Brock-Lalonishok-Lebaron-1992-JF} to test the profitability of our technical trading rules. The procedure are as follows,

\begin{enumerate}
\item For a given trading day, the possible position state can be long, short, or closed. For each trading rule, we assign one of the three position states to each trading day. Once the position state is determined, we can estimate the mean daily return condition on the long positions $\mu_l$ and the short positions $\mu_s$. We also calculate the mean daily return over the whole sample  $\mu$, which can be understood as the return obtained from the buy-and-hold strategy.

\item $t$-test is adopted to check whether mean daily return conditioned on the trading rules ($\mu_l$ and $\mu_s$)  is significantly different from the mean daily return over the whole sample $\mu$. The null hypothesis of our test is that mean conditional daily returns are equal to mean unconditional daily returns. The statistics of $t$-test is defined as follows,
\begin{equation}
\label{Eq:Tstatistics}
t_l=\frac{\mu_l-\mu}{\sqrt{\sigma^2/N_l+\sigma^2/N}}, ~~~~ t_s=\frac{\mu_s-\mu}{\sqrt{\sigma^2/N_s+\sigma^2/N}} ,
\end{equation}
where $N_l$ and $N_s$ are the number of days in long and short positions, and $N$ is the number of observations. $\sigma^2$ is the variance of the return over the entire sample. The evidence that technical trading rules have predictive ability will be supported if the mean daily returns in long (short) positions are positive (negative) and also significantly different from the mean daily returns $\mu$.

\item We further use the $t$-test to check whether the mean daily returns of long positions $\mu_l$ are significantly different from the mean daily returns of short positions $\mu_s$. The null hypothesis of our tests is that $\mu_l$ equals to $\mu_s$. The  statistics $t_{ls}$ is defined as follows,
\begin{equation}
\label{Equation:t1 test}
t_{ls}=\frac{\mu_l-\mu_s}{\sqrt{\sigma^2/N_l+\sigma^2/N_s}} \,\, .
\end{equation}
The result that $\mu_l$ is statistically significantly different from $\mu_s$ indicates that our technical trading rules are useful.

\item The Sharpe ratio, which measures the average excess return per unit of total risk, is calculated for both long and short positions. The formulae of Sharpe ratios ($s_l$ and $s_s$) are as follows,
\begin{equation}
s_l =\frac{\mu_l-r_f}{\sigma_l}, ~~~~ s_s =\frac{\mu_s-r_f}{\sigma_s},
\end{equation}
 where $r_f$ is the mean daily risk-free during the whole period, $\sigma_l$ and $\sigma_s$ are the variances of the return sample in long and short positions. Note that the larger the Sharpe ratio is, the better the trading rule is.
\end{enumerate}

\subsubsection{White's Reality Check Method}

Data snooping will occur when one uses a financial time series more than once for the purpose of inference or model selection. We here employ the White's Reality Check (WRC) test \citep{White-2000-Em} to  correct for the data snooping effects, which allows us to check whether the profitability of our technical trading rules is truly from the rules, or from pure luck. The null hypothesis of our WRC test is that the best trading rules in our strategy pools have no predictive superiority over a given benchmark strategy. Rejection of this null hypothesis implies that the best technical trading rules achieve performance superior to the benchmark. The procedure for the WRC test is as follows,

The performance statistic for trading rule $k$ is defined as follows,
\begin{equation}\label{Eq:fk}
\overline{f_k}=\frac{1}{n}\sum_{t=R}^{T}(f_{k, t+1}),
\end{equation}
where $f_{k, t+1}$ represents the comparison of the trading rule $k$ and benchmark from day $R$ to day $T$, and $n = T-R+1$. Since some trading rules require 200 days to generate trading signals, we will set $R = 201$. We also have $T = 5276$ and $5593$ for SHCI and SZCI respectively. The performance measure $f_{k, t+1}$ is written as,
\begin{equation}
\label{Eq:fkt}
f_{k,t+1}= \log(1 + y_{t+1} I_{k, t+1})- \log(1 + y_{t+1} I_{0, t+1})   \qquad k=1,\ldots,48
\end{equation}
where $y_{t}$ is the relative return on day $t$, defined as $y_{t}= (p_{t} - p_{t-1}) / p_{t-1}$. $I_{k, t}$ and $I_{0, t}$ are the market positions on day $t$ which are converted from the trading signals generated from the trading rule $k$ and the benchmark. If sell short mechanism is allowed, this dummy variable $I$ will take one of the three values: $I=1$ represents a long position, $I=-1$ represents a short position, and $I=0$ represents a neutral position. If sell short mechanism is not allowed, the dummy variable $I$ will take one of the two values: long positions are represented by $I=1$ and other positions are represented by $I=0$. We also set $I_0=1$ to stand for the buy-and-hold strategy.

In order to make our performance measure closer to the actual situation, the transaction costs would be included in the measure $f_{k, t+1}$,
\begin{equation}
\label{Equation:fktc}
f_{k,t+1}=\log(1+y_{t+1} I_{k, t+1} - c|I_{k,t+1} - I_{k,t}|)- \log(1 + y_{t+1} I_{0, t+1})   \qquad k=1, \ldots, 48
\end{equation}
where $c$ stands for one-way transaction cost rate. In the Chinese stock market, the transaction costs mainly consist of stamp duty, commission and transfer fees. Stamp duty is levied on the transferor, which is 0.1\% of the turnover since September 19, 2008. Commission is levied bidirectionally, the maximum amount of which is no more than 0.3\% of the turnover. Transfer fee is charged only in the Shanghai market, which is 0.06\% of the denomination. To simplify the process of estimating transaction costs, we use four rates ($c=0, 0.3\%, 0.5\%$, and $1\%$) to include all the stamp duty, commission, transfer fees and other costs.

Based on our performance measurement, we formulate the null hypothesis of WRC tests as follows,
\begin{equation}
\label{Eq:H0}
H_0: \max_{k=1,\ldots,48}{E(f_k)} \leq 0 .
\end{equation}
We evaluate this hypothesis $H_0$ by applying the stationary bootstrap method \citep{Politis-Romano-1994-JASA} on values of $f_{k, t}$ for each trading rule \citep{White-2000-Em}. In the stationary bootstrap method, the synthetic data $f^*_{k,t}$ for rule $k$ are obtained from block shuffling the performance series $f_{k,t}$. The size of the blocks is determined by a predefined ``smoothing parameter'' $q$, which gives the expected length of the shuffled block as $1/q$. In our tests, we use four different values of the smoothing parameter ($q=0.01, 0.1, 0.5$ and $1$) to check whether the smoothing parameter has effects on WRC tests. Values of $q$ correspond to mean block lengths of 100, 10, 2 and 1 respectively. For a given series $\theta(t)$ with $R \le t \le T$, the block resampling series $\theta^*(t)$ can be obtained through the following procedure,

\begin{enumerate}
\item Set $t=R$ and $\theta^*(t)=\theta(i)$, where $i$ is random, independently and uniformly drawn from $R,\ldots,T$.
\item Increasing $t$ by 1. If $t>T$, stop. Otherwise, draw $u$ from the standard uniform distribution $[0,1]$. If $u<q$, set $\theta^*(t) = \theta(i)$, where $i$ is random, independently and uniformly  drawn from $R,\ldots,T$. If $u\geq q$, set $\theta^*(t)=\theta(i+1)$; if $i+1 > T$, we reset $i=R$.
\item Repeat step two.
\end{enumerate}

For trading rule $k$, the performance statistics of synthetic data $\overline{f^*_k}$ can be evaluated from Eq.~(\ref{Eq:fk}). We construct the following statistics to obtain the $p$-value,
\begin{equation}
\label{Eq:V}
\overline{V}=\max_{k=1,\ldots,48}{\sqrt{n}(\overline{f_k})}  \,\,\, ,
\end{equation}
\begin{equation}
\label{Eq:V1}
\overline{V^{*}}=\max_{k=1,\ldots,48}{\sqrt{n}(\overline{f^*_k}-\overline{f_k})} \,\, .
\end{equation}
We accumulate 500 values of $\overline{V^*}$ and estimate the $p$-value as,
\begin{equation}
\label{Eq:pvalue}
{\rm{Pr}}(\overline{V^*} > \overline{V})  .
\end{equation}
If the $p$ value is smaller than a certain significance level, the null hypothesis is rejected. The best technical trading rules can outperform the buy-and-hold strategy accounting for data snooping effects.

\section{Empirical results}
\subsection{Results of traditional $t$-tests}

In order to check the validity of our trading strategies based on VMA, FMA and TRB rules, we perform a back test on two the indexes, SHCI and SZCI. The results are shown in Table~\ref{T:SHCI} (SHCI) and \ref{T:SZCI} (SZCI). The first column of the two tables lists the trading rules with their corresponding parameters. The number of days in long positions $N_l$ and short positions $N_s$ are reported in the column 2 and 3. We notice that the increment of short term window size $n_s$ and long term window size $n_l$ will reduce the number of days holding for long and short positions. The larger the window size is, the smoother the moving average line is, which will reduce the number of trading signals. We also find that the number of days in long and short positions decreases when we take a band into consideration, except for FMA rules with $n_l=50$. This implies that the band does have the ability to remove the noisy signals in turbulent market.

\setlength\tabcolsep{2pt}
\begin{table*}[htbp]
\centering \small
 \caption{\label{T:SHCI} Results of our trading strategies on SHCI. $N$ represent the number of days in the same market positions. The returns $\mu$ have been multiplied by a factor of $10^4$. Standard deviations $\sigma$ and Sharpe ratios $s$ have been multiplied by a factor of $10^2$. $p$ are the fraction of returns on signals higher than zero. The subscripts $l$ and $s$ represent long and short positions respectively. The superscripts $*$, $**$, and $***$ represent the significance level of 10\%, 5\%, and 1\%.}
 \medskip
 \centering
\begin{tabular}{lrrcr@{.}lr@{.}lr@{.}lcccccccr@{.}lr@{.}l}
 \toprule
  {Trading rule} & \multicolumn{2}{c}{$N$} && \multicolumn{6}{c}{$\mu$} && \multicolumn{2}{c}{$\sigma$} && \multicolumn{2}{c}{$p$} && \multicolumn{4}{c}{$s$} \\
    \cline{2-3} \cline{5-10}  \cline{12-13}  \cline{15-16}  \cline{18-21}
   parameters  & \multicolumn{1}{c}{$N_l$}    &  \multicolumn{1}{c}{$N_s$}    && \multicolumn{2}{c}{$\mu_l$} & \multicolumn{2}{c}{$\mu_s$} & \multicolumn{2}{c}{$\Delta \mu$} && $\sigma_l$  & $\sigma_s$  && $p_l$ & $p_s$ && \multicolumn{2}{c}{$s_l$} & \multicolumn{2}{c}{$s_s$} \\
 \hline
    VMA $(1, 20, 0)$     & 2525  & 2551  && {\bf{13}}&{\bf{17}}$^{**}$  & {\bf{$-$10}}&{\bf{92}}$^{**}$  & {\bf{24}}&{\bf{10}}$^{***}$ && 2.00  & 2.23  && 0.56  & 0.47  && 6&11  & $-$5&32  \\
    VMA $(2, 20, 0)$     & 2520  & 2556  && {\bf{14}}&{\bf{03}}$^{**}$ & {\bf{$-$11}}&{\bf{72}}$^{**}$  & {\bf{25}}&{\bf{75}}$^{***}$ && 2.00  & 2.23  && 0.56  & 0.47  && 6&54  & $-$5&68  \\
    VMA $(5, 20, 0)$     & 2522  & 2554  && 8&33  & $-$6&10  & {\bf{14}}&{\bf{43}}$^{**}$ && 1.98  & 2.25  && 0.55  & 0.48  && 3&73  & $-$3&12  \\
    VMA $(1, 50, 0)$     & 2468  & 2608  &&  8&61  & $-$6&07  & {\bf{14}}&{\bf{68}}$^{**}$ && 2.00  & 2.23  && 0.56  & 0.47  && 3&84  & $-$3&14  \\
    VMA $(2, 50, 0)$     & 2469  & 2607  && 7&77  & $-$5&28  & {\bf{13}}&{\bf{05}}$^{**}$ && 2.01  & 2.23  && 0.56  & 0.47  && 3&40  & $-$2&79  \\
    VMA $(5, 50, 0)$     & 2468  & 2608  && 6&25  & $-$3&84  & {\bf{10}}&{\bf{10}}$^{*}$ && 2.00  & 2.23  && 0.55  & 0.48  && 2&66  & $-$2&14  \\
    VMA $(1, 150, 0)$     & 2602  & 2474  && 3&51  & $-$1&50  & 5&00  && 2.11  & 2.14  && 0.54  & 0.49  && 1&22  & $-$1&14  \\
    VMA $(2, 150, 0)$     & 2601  & 2475  && 4&28  & $-$2&31  & 6&58  && 2.14  & 2.11  && 0.54  & 0.49  && 1&56  & $-$1&54  \\
    VMA $(5, 150, 0)$     & 2612  & 2464  && 3&04  & $-$1&03  & 4&07  && 2.10  & 2.15  && 0.54  & 0.49  && 1&00  & $-$0&91  \\
    VMA $(1, 200, 0)$     & 2544  & 2532  && 1&62  & 0&51  & 1&10  && 2.11  & 2.14  && 0.54  & 0.49  && 0&32  & $-$0&20  \\
    VMA $(2, 200, 0)$     & 2543  & 2533  && 3&42  & $-$1&30  & 4&72  && 2.14  & 2.10  && 0.54  & 0.49  && 1&16  & $-$1&06  \\
    VMA $(5, 200, 0)$     & 2546  & 2530  && 2&00  & 0&13  & 1&86  && 2.11  & 2.14  && 0.54  & 0.49  && 0&50  & $-$0&38  \\
    \hline
    VMA $(1, 20, 0.01)$ & 2107  & 2152  && {\bf{15}}&{\bf{47}}$^{***}$ & {\bf{$-$10}}&{\bf{67}}$^{**}$  & {\bf{26}}&{\bf{13}}$^{***}$ && 2.11  & 2.35  && 0.56  & 0.47  && 6&90  & $-$4&94  \\
    VMA $(2, 20, 0.01)$  & 2090  & 2127  && {\bf{16}}&{\bf{72}}$^{***}$ & {\bf{$-$9}}&{\bf{84}}$^{**}$  & {\bf{26}}&{\bf{57}}$^{***}$ && 2.10  & 2.35  && 0.57  & 0.48  && 7&52  & $-$4&58  \\
    VMA $(5, 20, 0.01)$  & 1986  & 2007  && {\bf{11}}&{\bf{50}}$^{*}$ & $-$6&77  & {\bf{18}}&{\bf{27}}$^{***}$ && 2.12  & 2.43  && 0.56  & 0.48  && 4&98  & $-$3&17  \\
    VMA $(1, 50, 0.01)$  & 2232  & 2356  && 9&17  & $-$3&82  & {\bf{12}}&{\bf{99}}$^{**}$ && 2.06  & 2.28  && 0.56  & 0.48  && 3&99  & $-$2&09  \\
    VMA $(2, 50, 0.01)$  & 2221  & 2359  && 9&62  & $-$5&36  & {\bf{14}}&{\bf{98}}$^{**}$ && 2.05  & 2.28  && 0.56  & 0.47  && 4&23  & $-$2&76  \\
    VMA $(5, 50, 0.01)$  & 2225  & 2339  && 6&54  & $-$4&06  & {\bf{10}}&{\bf{60}}$^{*}$ && 2.04  & 2.31  && 0.56  & 0.47  && 2&75  & $-$2&16  \\
    VMA $(1, 150, 0.01)$  & 2436  & 2324  && 3&82  & $-$2&59  & 6&41  && 2.14  & 2.18  && 0.54  & 0.49  && 1&35  & $-$1&62  \\
    VMA $(2, 150, 0.01)$  & 2447  & 2317  && 4&80  & $-$1&97  & 6&77  && 2.17  & 2.13  && 0.54  & 0.49  && 1&78  & $-$1&37  \\
    VMA $(5, 150, 0.01)$  & 2437  & 2306  && 3&60  & $-$1&28  & 4&87  && 2.15  & 2.20  && 0.54  & 0.49  && 1&24  & $-$1&01  \\
    VMA $(1, 200, 0.01)$  & 2388  & 2398  && 2&09  & 0&54  & 1&55  && 2.14  & 2.15  && 0.54  & 0.49  && 0&54  & $-$0&18  \\
    VMA $(2, 200, 0.01)$  & 2396  & 2393  && 1&99  & $-$0&93  & 2&92  && 2.11  & 2.14  && 0.54  & 0.49  && 0&50  & $-$0&87  \\
    VMA $(5, 200, 0.01)$  & 2400  & 2375  && 1&32  & $-$0&27  & 1&59  && 2.14  & 2.17  && 0.54  & 0.49  && 0&18  & $-$0&56  \\
    \hline
    FMA $(1, 50, 0)$     & 610   & 580   && 14&01  & $-$13&07  & {\bf{27}}&{\bf{08}}$^{**}$  && 2.25  & 2.02  && 0.56  & 0.47  && 5&82  & $-$6&93  \\
    FMA $(2, 50, 0)$     & 530   & 600   && {\bf{18}}&{\bf{01}}$^{*}$  & $-$11&60  & {\bf{29}}&{\bf{60}}$^{**}$ && 2.36  & 1.98  && 0.58  & 0.48  && 7&23  & $-$6&32  \\
    FMA $(5, 50, 0)$     & 510   & 560   && {\bf{18}}&{\bf{21}}$^{*}$  & {\bf{$-$15}}&{\bf{35}}$^{*}$  & {\bf{33}}&{\bf{56}}$^{**}$  && 2.10  & 1.99  && 0.60  & 0.48  && 8&21  & $-$8&20  \\
    FMA $(1, 150, 0)$     & 310   & 410   && 6&80  & 4&68  & 2&12  && 2.54  & 2.37  && 0.53  & 0.51  && 2&30  & 1&58  \\
    FMA $(2, 150, 0)$     & 310   & 330   && 0&45  & $-$2&79  & 3&24  && 2.49  & 2.44  && 0.52  & 0.49  && $-$0&20  & $-$1&53  \\
    FMA $(5, 150,  0)$     & 270   & 290   && $-$5&40  & $-$4&41  & $-$0&99  && 2.18  & 2.11  && 0.51  & 0.47  && $-$2&91  & $-$2&54  \\
    FMA $(1, 200, 0)$     & 310   & 370   && 4&65  & $-$0&54  & 5&19  && 2.80  & 2.33  && 0.51  & 0.49  && 1&32  & $-$0&63  \\
    FMA $(2, 200, 0)$     & 290   & 300   && $-$6&73  & $-$7&08  & 0&36  && 2.69  & 2.21  && 0.49  & 0.47  && $-$2&84  & $-$3&64  \\
    FMA $(5, 200, 0)$     & 270   & 290   && $-$8&73  & 2&49  & $-$11&22  && 2.27  & 2.11  && 0.50  & 0.49  && $-$4&26  & 0&74  \\
    \hline
    FMA $(1, 50, 0.01)$  & 590   & 560   && {\bf{16}}&{\bf{29}}$^{*}$  & $-$13&15  & {\bf{29}}&{\bf{44}}$^{**}$  && 2.24  & 2.06  && 0.56  & 0.47  && 6&84  & $-$6&83  \\
    FMA $(2, 50, 0.01)$  & 520   & 540   && 16&39  & $-$10&82  & {\bf{27}}&{\bf{21}}$^{**}$  && 2.26  & 2.12  && 0.56  & 0.48  && 6&85  & $-$5&56  \\
    FMA $(5, 50, 0.01)$  & 510   & 500   && 12&26  & {\bf{$-$21}}&{\bf{63}}$^{**}$  & {\bf{33}}&{\bf{89}}$^{**}$  && 2.01  & 2.06  && 0.56  & 0.45  && 5&64  & $-$10&93  \\
    FMA $(1, 150, 0.01)$  & 320   & 370   && 17&43  & $-$6&21  & 23&64  && 2.66  & 2.24  && 0.54  & 0.49  && 6&20  & $-$3&19  \\
    FMA $(2, 150, 0.01)$  & 280   & 340   && 10&16  & $-$5&30  & 15&46  && 2.55  & 2.30  && 0.55  & 0.47  && 3&63  & $-$2&71  \\
    FMA $(5, 150, 0.01)$  & 200   & 300   && 0&84  & $-$14&79  & 15&63  && 2.43  & 2.08  && 0.55  & 0.46  && $-$0&04  & $-$7&57  \\
    FMA $(1, 200, 0.01)$  & 300   & 340   && 0&95  & 5&64  & $-$4&69  && 2.95  & 2.13  && 0.50  & 0.51  && 0&01  & 2&21  \\
    FMA $(2, 200, 0.01)$  & 280   & 300   && $-$3&75  & $-$2&70  & $-$1&05  && 2.38  & 2.42  && 0.51  & 0.49  && $-$1&97  & $-$1&50  \\
    FMA $(5, 200, 0.01)$  & 310   & 210   && $-$13&62  & 1&21  & $-$14&84  && 2.21  & 2.31  && 0.50  & 0.51  && $-$6&59  & 0&12  \\
    \hline
    TRB $(50, 0)$     & 1150  & 1110  && 11&78  & $-$5&26  & {\bf{17}}&{\bf{04}}$^{*}$  && 2.16  & 2.01  && 0.56  & 0.47  && 5&01  & $-$3&09  \\
    TRB $(150, 0)$   & 690   & 600   && {\bf{17}}&{\bf{85}}$^{*}$  & $-$4&56  & {\bf{22}}&{\bf{41}}$^{*}$  && 2.10  & 2.18  && 0.59  & 0.47  && 8&06  & $-$2&52  \\
    TRB $(200, 0)$ & 600   & 510   && 15&07  & 1&59  & 13&47  && 2.01  & 2.29  && 0.60  & 0.48  && 7&01  & 0&29  \\
    \hline
    TRB $(50, 0.01)$ & 730   & 870   && {\bf{20}}&{\bf{04}}$^{**}$  & $-$5&84  & {\bf{25}}&{\bf{88}}$^{**}$  && 2.48  & 2.14  && 0.59  & 0.47  && 7&70  & $-$3&16  \\
    TRB $(150, 0.01)$  & 460   & 430   && {\bf{21}}&{\bf{46}}$^{**}$  & $-$7&69  & {\bf{29}}&{\bf{16}}$^{**}$  && 2.39  & 2.32  && 0.61  & 0.47  && 8&59  & $-$3&72  \\
    TRB $(200, 0.01)$  & 400   & 390   && 18&06  & $-$1&32  & 19&38  && 2.24  & 2.38  && 0.61  & 0.48  && 7&64  & $-$0&95  \\
\bottomrule
\end{tabular}
\end{table*}

\setlength\tabcolsep{2pt}
\begin{table*}[htbp]
\centering \small
 \caption{\label{T:SZCI} Results of our trading strategies on SZCI. $N$ represent the number of days in the same market positions. The returns $\mu$ have been multiplied by a factor of $10^4$. Standard deviations $\sigma$ and Sharpe ratios $s$ have been multiplied by a factor of $10^2$. $p$ are the fraction of returns on signals higher than zero. The subscripts $l$ and $s$ represent long and short positions respectively. The superscripts $*$, $**$, and $***$ represent the significance level of 10\%, 5\%, and 1\%.}
 \medskip
 \centering
\begin{tabular}{lrrcr@{.}lr@{.}lr@{.}lcccccccr@{.}lr@{.}l}
 \toprule
  {Trading rule} & \multicolumn{2}{c}{$N$} && \multicolumn{6}{c}{$\mu$} && \multicolumn{2}{c}{$\sigma$} && \multicolumn{2}{c}{$p$} && \multicolumn{4}{c}{$s$} \\
    \cline{2-3} \cline{5-10}  \cline{12-13}  \cline{15-16}  \cline{18-21}
   parameters  & \multicolumn{1}{c}{$N_l$}    &  \multicolumn{1}{c}{$N_s$}    && \multicolumn{2}{c}{$\mu_l$} & \multicolumn{2}{c}{$\mu_s$} & \multicolumn{2}{c}{$\Delta \mu$} && $\sigma_l$  & $\sigma_s$  && $p_l$ & $p_s$ && \multicolumn{2}{c}{$s_l$} & \multicolumn{2}{c}{$s_s$} \\
 \hline
    VMA $(1, 20, 0)$     & 2769  & 2624  && {\bf{19}}&{\bf{42}}$^{***}$ & {\bf{$-$11}}&{\bf{75}}$^{***}$  & {\bf{31}}&{\bf{17}}$^{***}$ && 2.07  & 2.31  && 0.58  & 0.48  && 8&93  & $-$5&50  \\
    VMA $(2, 20, 0)$     & 2752  & 2641  && {\bf{19}}&{\bf{96}}$^{***}$ & {\bf{$-$12}}&{\bf{12}}$^{***}$  & {\bf{32}}&{\bf{08}}$^{***}$ && 2.10  & 2.28  && 0.58  & 0.47  && 9&09  & $-$5&72  \\
    VMA $(5, 20, 0)$     & 2746  & 2647  && {\bf{15}}&{\bf{76}}$^{**}$ & {\bf{$-$7}}&{\bf{69}}$^{**}$  & {\bf{23}}&{\bf{45}}$^{***}$ && 2.10  & 2.28  && 0.57  & 0.48  && 7&06  & $-$3&77  \\
    VMA $(1, 50, 0)$     & 2777  & 2616  && {\bf{14}}&{\bf{02}}$^{*}$ & {\bf{$-$6}}&{\bf{12}}$^{**}$  & {\bf{20}}&{\bf{14}}$^{***}$ && 2.13  & 2.26  && 0.57 & 0.48  && 6&15  & $-$3&12  \\
    VMA $(2, 50, 0)$    & 2769  & 2624  && {\bf{13}}&{\bf{31}}$^{*}$ & {\bf{$-$5}}&{\bf{31}}$^{*}$  & {\bf{18}}&{\bf{62}}$^{***}$ && 2.14  & 2.24  && 0.57  & 0.48  && 5&78  & $-$2&78  \\
    VMA $(5, 50, 0)$     & 2770  & 2623  && 12&06  & $-$3&99  & {\bf{16}}&{\bf{05}}$^{***}$ && 2.13  & 2.26  && 0.57  & 0.48  && 5&24  & $-$2&18  \\
    VMA $(1, 150, 0)$     & 2758  & 2635  && 11&27  & $-$3&10  & {\bf{14}}&{\bf{37}}$^{**}$ && 2.20  & 2.18  && 0.57  & 0.49  && 4&70  & $-$1&84  \\
    VMA $(2, 150, 0)$     & 2760  & 2633  && 12&04  & $-$3&91  & {\bf{15}}&{\bf{95}}$^{***}$ && 2.19  & 2.20  && 0.57  & 0.48  && 5&07  & $-$2&20  \\
    VMA $(5, 150, 0)$     & 2743  & 2650  && 10&94  & $-$2&67  & {\bf{13}}&{\bf{60}}$^{**}$ && 2.18  & 2.21  && 0.57  & 0.49  && 4&59  & $-$1&63  \\
    VMA $(1, 200, 0)$     & 2690  & 2703  && 11&73  & $-$3&19  & {\bf{14}}&{\bf{91}}$^{**}$ && 2.18  & 2.20  && 0.57  & 0.48  && 4&94  & $-$1&87  \\
    VMA $(2, 200, 0)$     & 2692  & 2701  && 10&92  & $-$2&39  & {\bf{13}}&{\bf{30}}$^{**}$ && 2.18  & 2.21  && 0.57  & 0.48  && 4&58  & $-$1&50  \\
    VMA $(5, 200, 0)$     & 2699  & 2694  && 11&42  & $-$2&93  & {\bf{14}}&{\bf{35}}$^{**}$ && 2.17  & 2.22  && 0.57  & 0.48  && 4&83  & $-$1&74  \\
    \hline
    VMA $(1, 20, 0.01)$  & 2320  & 2206  && {\bf{23}}&{\bf{50}}$^{***}$ & {\bf{$-$10}}&{\bf{92}}$^{***}$  & {\bf{34}}&{\bf{42}}$^{***}$ && 2.15  & 2.43  && 0.59  & 0.48  && 10&50  & $-$4&87  \\
    VMA $(2, 20, 0.01)$  & 2282  & 2185  && {\bf{23}}&{\bf{58}}$^{***}$ & {\bf{$-$10}}&{\bf{22}}$^{***}$  & {\bf{33}}&{\bf{79}}$^{***}$ && 2.19  & 2.40  && 0.59  & 0.48  && 10&35  & $-$4&64  \\
    VMA $(5, 20, 0.01)$  & 2176  & 2124  && {\bf{21}}&{\bf{34}}$^{***}$ & {\bf{$-$6}}&{\bf{80}}$^{**}$  & {\bf{28}}&{\bf{14}}$^{***}$ && 2.20  & 2.43  && 0.59  & 0.48  && 9&27  & $-$3&18  \\
    VMA $(1, 50, 0.01)$  & 2488  & 2368  && {\bf{16}}&{\bf{80}}$^{**}$ & {\bf{$-$6}}&{\bf{90}}$^{**}$  & {\bf{23}}&{\bf{70}}$^{***}$ && 2.19  & 2.31  && 0.58  & 0.48  && 7&24  & $-$3&38  \\
    VMA $(2, 50, 0.01)$  & 2475  & 2361  && {\bf{15}}&{\bf{47}}$^{**}$ & {\bf{$-$7}}&{\bf{31}}$^{**}$  & {\bf{22}}&{\bf{77}}$^{***}$ && 2.20  & 2.30  && 0.58  & 0.48  && 6&59  & $-$3&58  \\
    VMA $(5, 50, 0.01)$  & 2468  & 2373  && {\bf{13}}&{\bf{80}}$^{*}$ & $-$3&08  & {\bf{16}}&{\bf{88}}$^{***}$ && 2.20  & 2.33  && 0.58  & 0.48  && 5&85  & $-$1&72  \\
    VMA $(1, 150, 0.01)$  & 2607  & 2478  && 11&75  & $-$2&59  & {\bf{14}}&{\bf{35}}$^{**}$ && 2.22  & 2.22  && 0.57  & 0.49  && 4&86  & $-$1&59  \\
    VMA $(2, 150, 0.01)$  & 2608  & 2476  && 12&51  & $-$2&94  & {\bf{15}}&{\bf{45}}$^{**}$ && 2.22  & 2.23  && 0.57  & 0.48  && 5&21  & $-$1&74  \\
    VMA $(5, 150, 0.01)$  & 2615  & 2469  && 12&20  & $-$2&33  & {\bf{14}}&{\bf{53}}$^{**}$ && 2.21  & 2.25  && 0.57  & 0.49  && 5&09  & $-$1&45  \\
    VMA $(1, 200,  0.01)$  & 2605  & 2606  && 12&64  & $-$3&14  & {\bf{15}}&{\bf{79}}$^{***}$ && 2.20  & 2.23  && 0.58  & 0.48  && 5&31  & $-$1&83  \\
    VMA $(2, 200, 0.01)$  & 2605  & 2597  && 11&92  & $-$2&82  & {\bf{14}}&{\bf{74}}$^{**}$ && 2.19  & 2.22  && 0.57  & 0.48  && 5&01  & $-$1&69  \\
    VMA $(5, 200, 0.01)$  & 2607  & 2595  && 11&36  & $-$3&28  & {\bf{14}}&{\bf{64}}$^{**}$ && 2.17  & 2.23  && 0.57  & 0.48  && 4&80  & $-$1&89  \\
    \hline
    FMA $(1, 50, 0)$  & 640   & 700   && 10&34  & $-$0&74  & 11&08  && 2.14  & 2.00  && 0.55  & 0.51  && 4&39  & $-$0&84  \\
    FMA $(2, 50, 0)$  & 620   & 660   && 13&83  & $-$2&18  & 16&01  && 2.17  & 2.07  && 0.56  & 0.49  && 5&94  & $-$1&50  \\
    FMA $(5, 50, 0)$  & 540   & 590   && 10&26  & $-$3&14  & 13&40  && 2.09  & 1.97  && 0.57  & 0.49  && 4&47  & $-$2&07  \\
    FMA $(1, 150, 0)$  & 390   & 310   && 1&88  & $-$2&11  & 3&98  && 2.50  & 2.10  && 0.50  & 0.50  && 0&38  & $-$1&45  \\
    FMA $(2, 150, 0)$  & 350   & 280   && 1&26  & $-$6&63  & 7&90  && 2.41  & 2.01  && 0.50  & 0.47  && 0&14  & $-$3&76  \\
    FMA $(5, 150, 0)$  & 290   & 260   && 19&40  & 0&95  & 18&45  && 2.71  & 1.96  && 0.51  & 0.50  && 6&82  & 0&01  \\
    FMA $(1, 200, 0)$  & 270   & 230   && 12&13  & 0&21  & 11&92  && 2.88  & 2.06  && 0.53  & 0.52  && 3&88  & $-$0&35  \\
    FMA $(2, 200, 0)$  & 250   & 220   && 3&62  & $-$8&65  & 12&27  && 2.90  & 2.27  && 0.52  & 0.51  && 0&93  & $-$4&21  \\
    FMA $(5, 200, 0)$  & 210   & 200   && 1&16  & $-$12&51  & 13&67  && 2.66  & 2.35  && 0.50  & 0.49  && 0&09  & $-$5&72  \\
    \hline
    FMA $(1, 50, 0.01)$  & 660   & 600   && 13&98  & $-$3&91  & 17&90  && 2.15  & 2.08  && 0.57  & 0.49  && 6&08  & $-$2&33  \\
    FMA $(2, 50, 0.01)$  & 660   & 550   && 12&41  & $-$7&39  & 19&80  && 2.23  & 1.99  && 0.56  & 0.47  && 5&15  & $-$4&18  \\
    FMA $(5, 50, 0.01)$  & 540   & 530   && 1&28  & 0&51  & 0&78  && 2.14  & 2.01  && 0.55  & 0.51  && 0&16  & $-$0&21  \\
    FMA $(1, 150, 0.01)$  & 380   & 300   && $-$0&64  & $-$9&59  & 8&95  && 2.40  & 2.13  && 0.49  & 0.47  && $-$0&65  & $-$4&94  \\
    FMA $(2, 150, 0.01)$  & 280   & 330   && 13&55  & $-$7&00  & 20&54  && 2.65  & 1.96  && 0.54  & 0.47  && 4&77  & $-$4&04  \\
    FMA $(5, 150, 0.01)$  & 230   & 310   && 19&77  & 7&61  & 12&15  && 2.52  & 2.44  && 0.54  & 0.49  && 7&48  & 2&73  \\
    FMA $(1, 200, 0.01)$  & 240   & 240   && 20&54  & $-$0&64  & 21&19  && 3.02  & 2.08  && 0.55  & 0.52  && 6&50  & $-$0&76  \\
    FMA $(2, 200, 0.01)$  & 230   & 210   && 13&23  & $-$4&12  & 17&35  && 3.02  & 2.22  && 0.53  & 0.50  && 4&08  & $-$2&28  \\
    FMA $(5, 200, 0.01)$  & 210   & 190   && 4&34  & $-$7&18  & 11&53  && 2.72  & 2.29  && 0.48  & 0.51  && 1&25  & $-$3&54  \\
    \hline
    TRB $(50, 0)$  & 1290  & 1220  && {\bf{18}}&{\bf{79}}$^{**}$  & $-$5&11  & {\bf{23}}&{\bf{90}}$^{***}$  && 2.49  & 2.19  && 0.60  & 0.50  && 7&16  & $-$2&76  \\
    TRB $(150, 0)$  & 850   & 620   && 16&02  & $-$0&80  & 16&82  && 2.45  & 2.45  && 0.60  & 0.49  && 6&17  & $-$0&71  \\
    TRB $(200, 0)$  & 780   & 490   && {\bf{20}}&{\bf{32}}$^{*}$  & $-$5&96  & {\bf{26}}&{\bf{27}}$^{**}$  && 2.33  & 2.46  && 0.60  & 0.47  && 8&32  & $-$2&81  \\
    \hline
    TRB $(50, 0.01)$  & 870   & 950   && {\bf{35}}&{\bf{91}}$^{***}$  & $-$6&86  & {\bf{42}}&{\bf{77}}$^{***}$  && 2.68  & 2.28  && 0.62  & 0.49  && 13&05  & $-$3&42  \\
    TRB $(150, 0.01)$  & 520   & 500   && {\bf{40}}&{\bf{92}}$^{***}$  & 5&28  & {\bf{35}}&{\bf{64}}$^{***}$  && 2.54  & 2.61  && 0.64  & 0.50  && 15&74  & 1&67  \\
    TRB $(200, 0.01)$  & 470   & 400   && {\bf{42}}&{\bf{23}}$^{***}$  & $-$3&49  & {\bf{45}}&{\bf{71}}$^{***}$  && 2.58  & 2.59  && 0.64  & 0.48  && 15&99  & $-$1&71  \\
\bottomrule
\end{tabular}
\end{table*}

The average returns in long positions $\mu_l$ are reported in column 4. One can find that 43 (respectively, 47) of 48 mean returns are positive for SHCI (respectively, SZCI). We also find that 11 (respectively, 16) of the 48 rules provide significant returns which reject the hypothesis of the long position returns $\mu_l$ equaling the unconditional return $\mu$ at the significance level of 0.1, for SHCI (respectively, SZCI). In general, the introduction of one percentage band will increase the corresponding returns for most of the trading rules. This further consolidates the fact that adding bands can reduce the noisy signals and increase the returns. Since short selling is not allowed in the Chinese stock market, the long position returns can represent the investment returns of our trading strategies. The trading rules TRB $(150, 0.01)$ (respectively, TRB $(200, 0.01)$) give the maximum average return at 0.002146 (respectively, 0.004223) for SHCI (respectively, SZCI).

The average returns in short positions $\mu_s$ are listed in column 5. We find that 40 (respectively, 43) of 48 mean returns are negative for SHCI (respectively, SZCI). There are 6 (respectively, 10) returns significantly different from the unconditional mean return $\mu$ at the significance level of 0.1 for SHCI (respectively, SZCI). We also notice that the TRB rules do not provide any significant $\mu_s$ for SHCI while only VMA rules give significant $\mu_s$ for SZCI. The minimum value of $\mu_s$ is $-0.002163$ (respectively, $-0.001275$) for SHCI (respectively, SZCI) , generated by FMA $(5, 50, 0.01)$ (respectively, VMA $(2,20,0)$).

The differences between average long position returns and average short position returns $\Delta \mu$ are reported in column 6. This quantity can be interpreted as the investment returns when short selling is allowed. First, we observe that 22 (respectively, 29) of 48 differences are significantly different from zero at the significance level of 0.1 for SHCI (respectively, SZCI), which means that the long position returns are significantly different from the short position returns. The FMA rules do not provide any significant differences for SZCI. The largest difference 0.003389 (respectively, 0.004571) for SHCI (SZCI) is generated by FMA $(5, 50, 0)$ (respectively, TRB $(200, 0.01)$). From the significant $\mu_l$, $\mu_s$ and $\mu$, we infer that our trading strategies have the ability to forecast market trends and earn excess returns in markets.

We also report the standard deviations ($\sigma_l$ and $\sigma_s$) of the returns in long and short positions in column 7 and 8. For VMA rules, one striking result is that $\sigma_s$ is greater than $\sigma_l$ when $n_l \le 50$ but the difference between the two is not significant when $n_l > 50$. This again suggests that the VMA rules with parameters $n_l \le 50$ are successful. Specifically, those rules give higher returns and lower volatilities for long positions but lower returns and higher volatilities for short positions. For FMA and TRB rules, the standard deviations exhibit the opposite behavior, that $\sigma_s$ is less than $\sigma_l$. For rules with the same window size, we find that the introduction of a band will increase the Sharpe ratio. We further estimate the average standard deviations of VMA, FMA and TRB rules for both return samples and find that $\langle \sigma_l^{\rm{VMA}} \rangle = 0.0208$, $\langle \sigma_s^{\rm{VMA}} \rangle = 0.0221$, $\langle \sigma_l^{\rm{FMA}} \rangle = 0.0241$, $\langle \sigma_s^{\rm{FMA}} \rangle = 0.0218$, $\langle \sigma_l^{\rm{TRB}} \rangle = 0.0223$, $\langle \sigma_s^{\rm{TRB}} \rangle = 0.0222$ for SHCI and $\langle \sigma_l^{\rm{VMA}} \rangle = 0.0217$, $\langle \sigma_s^{\rm{VMA}} \rangle = 0.0227$, $\langle \sigma_l^{\rm{FMA}} \rangle = 0.0252$, $\langle \sigma_s^{\rm{FMA}} \rangle = 0.0211$, $\langle \sigma_l^{\rm{TRB}} \rangle = 0.0251$, $\langle \sigma_s^{\rm{TRB}} \rangle = 0.0243$ for SZCI. From the results of both indexes, we observe that the standard deviation $\sigma$ in the Shanghai market is lower than that in the Shenzhen market for the same rule and the same position. This suggests that the SZCI is more volatile than the SHCI, although they share a very similar trend. Furthermore, we can find that the $\langle \sigma_l^{\rm{VMA}} \rangle < \langle \sigma_l^{\rm{TRB}} \rangle < \langle \sigma_l^{\rm{FMA}} \rangle$ for long positions and $\langle \sigma_s^{\rm{FMA}} \rangle < \langle \sigma_s^{\rm{VMA}} \rangle < \langle \sigma_s^{\rm{TRB}} \rangle$ for short positions.

The fractions of the returns with value greater than zero in the return samples conditioned on long and short positions are also listed in columns 9 and 10. Obviously, one can see that 93 of 96 $p_l$ are greater than 0.5 while 85 of 96 $p_s$ are less than 0.5 for both indexes. By calculating the average value of the fraction $p$ over VMA, FMA ,and TRB rules for both indexes, we obtain $\langle p_l^{\rm{VMA}} \rangle = 0.5487$, $\langle p_s^{\rm{VMA}} \rangle = 0.4824$, $\langle p_l^{\rm{FMA}} \rangle = 0.5354$, $\langle p_s^{\rm{FMA}} \rangle = 0.4822$, $\langle p_l^{\rm{TRB}} \rangle = 0.5932$, $\langle p_s^{\rm{TRB}} \rangle = 0.4733$ for SHCI and $\langle p_l^{\rm{VMA}} \rangle = 0.5752$, $\langle p_s^{\rm{VMA}} \rangle = 0.4823$, $\langle p_l^{\rm{FMA}} \rangle = 0.5301$, $\langle p_s^{\rm{FMA}} \rangle = 0.4954$, $\langle p_l^{\rm{TRB}} \rangle = 0.6176$, $\langle p_s^{\rm{TRB}} \rangle = 0.4858$ for SZCI. If our trading strategies are useless, the average fractions will be the same for both positions. However, our results do show that the two fractions are significantly different, which demonstrate the usefulness of our trading strategies.

We list the Sharpe ratios of our trading strategies in columns 11 and 12 to show the average excess return in per unit of total risk. One can see that the ratios decrease with the increment of the long term window size for VMA rules. For VMA and TRB rules, the introduction of a band will increase the Sharpe ratio. The largest ratio of $s_l = 8.59$ (respectively, $s_l = 15.99$) is given by TRB $(150, 0.01)$ (TRB $(200, 0.01)$) for SHCI (SZCI). These two trading rules also provide the largest long position returns. The smallest ratio of $s_s = -10.93$ (respectively, $s_s = -5.72$) is given by FMA $(5, 50, 0.01)$ (respectively, FMA $(5, 200, 0)$) for SHCI (SZCI). The FMA $(5, 50, 0.01)$ also gives the smallest short position returns for SHCI, while the FMA $(5, 200, 0)$ does not provide any significant returns. The average Sharpe ratios are $\langle s_l^{\rm{VMA}} \rangle = 0.0283$, $\langle s_s^{\rm{VMA}} \rangle = -0.0220$, $\langle s_l^{\rm{FMA}} \rangle = 0.0196$, $\langle s_s^{\rm{FMA}} \rangle = -0.0352$, $\langle s_l^{\rm{TRB}} \rangle = 0.0734$, $\langle s_s^{\rm{TRB}} \rangle = -0.0219$ for SHCI and $\langle s_l^{\rm{VMA}} \rangle = 0.0629$, $\langle s_s^{\rm{VMA}} \rangle = -0.0273$, $\langle s_l^{\rm{FMA}} \rangle = 0.0343$, $\langle s_s^{\rm{FMA}} \rangle = -0.0219$, $\langle s_l^{\rm{TRB}} \rangle = 0.1107$, $\langle s_s^{\rm{TRB}} \rangle = -0.0162$ for SZCI. One can observe that the Sharpe ratio in Shanghai market is lower than that in Shenzhen market except for the short position of VMA rules.  In general, the result from the average return in the Shanghai market is lower than the corresponding result in the Shenzhen market. We also find that the TRB rules give the largest Sharpe ratio in both markets for long positions, indicating that TRB strategies have greater predictive abilities than the MA strategies in the Chinese market. This result is in agreement with the findings of \cite{Bessembinder-Chan-1995-PBFJ}.

\subsection{Results of White's Reality Check}

The above analysis does provide an amazing picture that our trading strategies have the power of predicting market trends and some of the trading rules can generate high returns associated with high Sharpe ratios.  Due to the presence of the data snooping effect, we will need to check whether the profitability of our trading strategies is from the strategies or just from luck.  To do this, we here perform the WRC test on both indexes. The synthetic data of WRC test are generated by the block shuffling procedure, in which case the key parameter $q$ will determine the expected length of shuffled blocks. We choose $q$ to take  $0.01, 0.1, 0.5, 1$ respectively to check the influence of different $q$ values on the significance of WRC tests. Table \ref{T:WRC:pvalue} lists the results of WRC tests on SHCI and SZCI for different values of $q$. As shown in the first line of each panel, we find that all the $p$-values of our WRC tests are less than 0.1 under the condition of no transaction costs for all values of $q$.  Whether or not the short selling is taken into account, our trading strategies can forecast the market trend and earn excess returns over the buy-and-hold strategy for both indexes when there is no transaction costs.

\begin{table}[htbp]
  \centering
  \caption{Results of White's Reality Check on SHCI and SZCI with different costs for different values of $q$.} \label{T:WRC:pvalue}
    \begin{tabular}{ccccccc}
    \toprule
    \multirow{2}[2]{*}{Index} & {Short} &  \multirow{2}[2]{*}{cost}    & \multirow{2}[2]{*}{$q=0.01$} & \multirow{2}[2]{*}{$q=0.1$} & \multirow{2}[2]{*}{$q=0.5$} & \multirow{2}[2]{*}{$q=1$}   \\
    &selling&&&&& \\
    \hline
    \multirow{4}[2]{*}{SHCI} &\multirow{4}[2]{*}{No} &0   & \textbf{0.044} & \textbf{0.056} & \textbf{0.044} &  \textbf{0.066}\\
          & & 0.3\% & 0.104 & 0.128 & 0.108 & 0.136   \\
          & & 0.5\% & 0.176 & 0.188 & 0.170 & 0.208  \\
          & & 1\% & 0.444 & 0.458 & 0.428 & 0.466  \\
     \hline
     \multirow{4}[2]{*}{SZCI} & \multirow{4}[2]{*}{No} & 0   & \textbf{0.090} & \textbf{0.078} & \textbf{0.058} & \textbf{0.054} \\
           & &0.3\% & 0.166 & 0.134 & 0.102 & 0.114  \\
           & &0.5\% & 0.226 & 0.206 & 0.178 & 0.170  \\
           & &1\% & 0.434 & 0.426 & 0.400 & 0.448 \\
     \hline
     \multirow{4}[2]{*}{SHCI} &\multirow{4}[2]{*}{Yes}       &0    & \textbf{0.034} & \textbf{0.060} & \textbf{0.062} & \textbf{0.084} \\
             &&0.3\% &    0.150 & 0.148 & 0.160 & 0.178 \\
             &&0.5\% &   0.252 & 0.280 & 0.248 & 0.296 \\
             &&1\% &   0.650 & 0.648 & 0.620 & 0.650 \\
    \hline
      \multirow{4}[2]{*}{SZCI} & \multirow{4}[2]{*}{Yes}   & 0   & \textbf{0.040} & \textbf{0.060} & \textbf{0.058} & \textbf{0.050}  \\
           &&0.3\% & 0.136 & 0.140 & 0.152 & 0.160 \\
           &&0.5\% &  0.254 & 0.250 & 0.242 & 0.252 \\
           &&1\% &  0.632 & 0.564 & 0.566 & 0.616 \\
    \bottomrule
    \end{tabular}%
\end{table}%

In real market situations, the transaction cost, which cannot be ignored, is an important factor in trading strategy design. For simplicity, we choose the costs $c$ to be $0.3\%, 0.5\%, 1\%$ to check whether the induction of costs can change the significance of WRC tests. As shown in Table \ref{T:WRC:pvalue}, we find that none has passed the WRC tests at the significant level of 0.1 when the transaction costs are added to the performance measure. This indicates that our trading strategies are ineffective when there are transaction costs. Our results are in agreement with the efficient market hypothesis.

\begin{figure}[htbp]
\begin{center}
  \includegraphics[width=16cm]{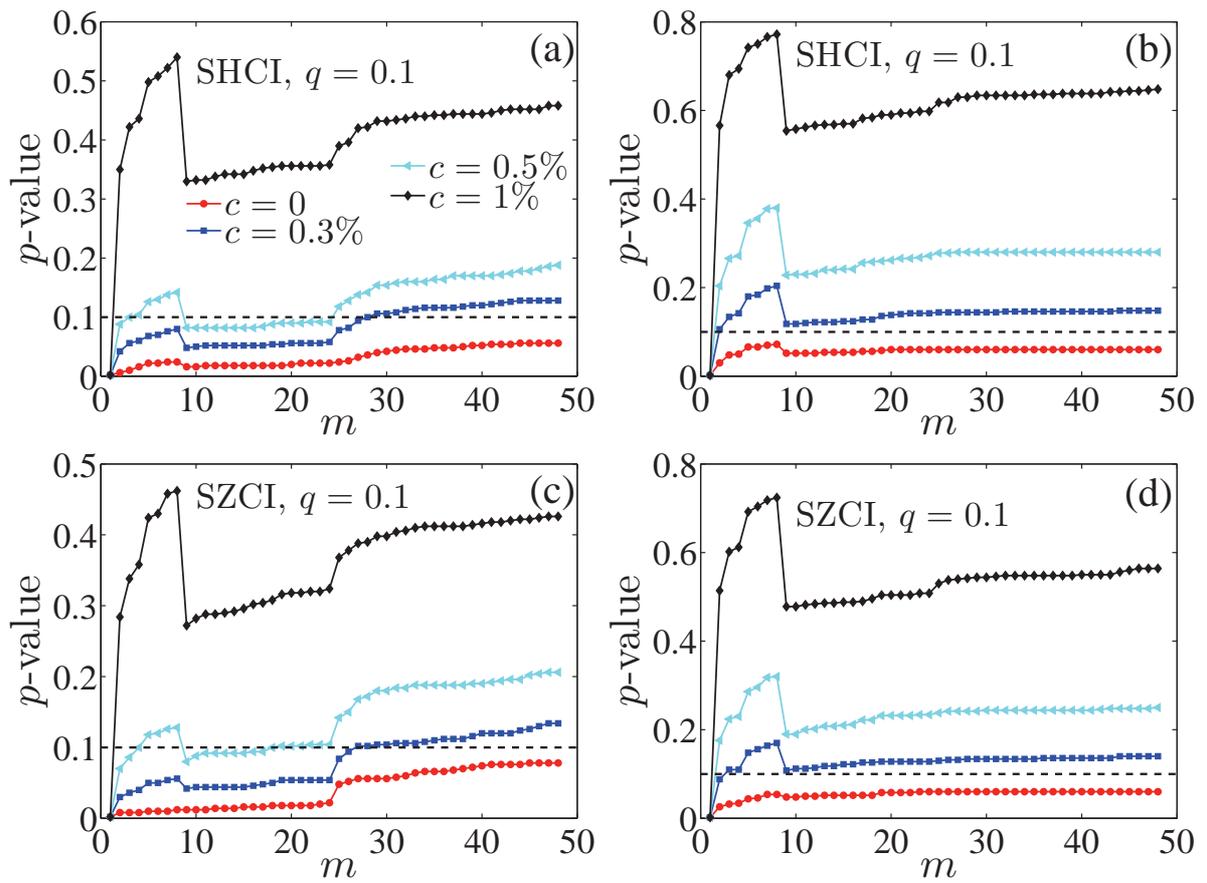}
  \caption{Results of White reality check on SHCI and SZCI with difference costs for $q=0.1$. (a, c) Short selling is not allowed. (b, d) Short selling is allowed.}
  \label{Fig:WRC}
\end{center}
\end{figure}

Figure \ref{Fig:WRC} illustrates the WRC's p-values as a function of the number of strategies $m$ with different transaction costs for both indexes. Both situations when the short selling is allowed or not are taken into consideration. We only show the results of $q=0.1$ here since the results of other $q$ values are similar. One can observe that only the red curves, which are associated with no transaction costs, in figure \ref{Fig:WRC} (a-d) are below the dash line ($p$-value = 0.1).

\section{Conclusion}

In this paper, we check the profitability of the technical trading rules, which consist of VMA, FMA, and TRB rules, by means of $t$-tests and WRC tests in Chinese stock markets. The $t$-tests indicate that VMA, FMA, and TRB rules are successful in forecasting stock price movements in the Shanghai market, whether or not short selling is allowed. Compared to the VMA and FMA rules, the TRB rules bring much higher returns.  Results from Shenzhen market are not coherent with the findings from the Shanghai market.  Only the trading stragtegies based on VMA and TRB rules produce useful trading signals. FMA rules however do not bring significant excess returns over buy-and-hold returns. The WRC tests indicate that the best trading rules significantly outperform the buy-and-hold strategy in both markets when there are no transaction costs, and whether or not the short selling is allowed. When transaction costs are taken into account, White's p-values show an upward tendency with the increase of transaction costs and the best trading rule no longer has superiority over the buy-and-hold strategy.

\section*{Acknowledgements}

This work was partially supported by the National Natural Science Foundation of China (11075054 and 71131007), Shanghai ``Chen Guang'' Project (2012CG34), Program for Changjiang Scholars and Innovative Research Team in University (IRT1028), and the Fundamental Research Funds for the Central Universities.

%\section{References}
%\bibliography{Bibliography}
%\bibliography{E:/Papers/Auxiliary/Bibliography}

%\end{CJK*}
\end{document}